\documentclass[12pt,twoside,a4paper,leqno]{article}
\usepackage[dvips,%
pdftitle={On a class of polynomial Lagrangians},%
pdfauthor={M. Palese, R. Vitolo},%
pdfkeywords={Fibred manifold, jet space, Lagrangian formalism,%
variational sequence}
pdfsubject={On a class of polynomial Lagrangians},%
colorlinks,linkcolor={blue},citecolor={blue},urlcolor={red}%
]{hyperref}
\usepackage{amsmath}
\usepackage{amssymb}
\usepackage[PostScript=dvips]{diagrams}
\newcommand*{\email}[1]{\href{mailto:#1}{\begingroup \urlstyle{rm}\Url{#1}}}
\newcommand*{\eprint}[1]{\href{http://arXiv.org/abs/#1}%
{\begingroup \Url{arXiv:#1}}}
\newtheorem{Def}{\indent Definition}[section]
\newtheorem{Lem}[Def]{\indent Lemma}
\newtheorem{Prop}[Def]{\indent Proposition}
\newtheorem{Theo}[Def]{\indent Theorem}
\newtheorem{Cor}[Def]{\indent Corollary}
\newtheorem{Rem}[Def]{\indent Remark}
\newtheorem{Exa}[Def]{\indent Example}
\newtheorem{Pro}[Def]{\indent Problem}
\newcommand{\bDf}{\begin{Def}\em}
\newcommand{\eDf}{\end{Def}}
\newcommand{\bLm}{\begin{Lem}}
\newcommand{\eLm}{\end{Lem}}
\newcommand{\bPr}{\begin{Prop}}
\newcommand{\ePr}{\end{Prop}}
\newcommand{\bTh}{\begin{Theo}}
\newcommand{\eTh}{\end{Theo}}
\newcommand{\bCr}{\begin{Cor}}
\newcommand{\eCr}{\end{Cor}}
\newcommand{\bRm}{\begin{Rem}\em}
\newcommand{\eRm}{\end{Rem}}
\newcommand{\bEx}{\begin{Exa}\em}
\newcommand{\eEx}{\end{Exa}}
\newcommand{\bPb}{\begin{Pro}\em}
\newcommand{\ePb}{\end{Pro}}
\newcommand{\edg}{\end{diagram}}
\newcommand{\bdg}{\begin{diagram}}

\newcommand{\R}{\mathbb{R}}

\newcommand{\bEq}{\begin{eqnarray}}
\newcommand{\eEq}{\end{eqnarray}}
\newcommand{\beq}{\begin{eqnarray*}}
\newcommand{\eeq}{\end{eqnarray*}}
\newcommand{\bPf}{\par\vspace*{-4pt}\indent{\sc Proof.}\enskip}
\newcommand{\ePf}{\medskip}
\def\QED{\hskip0.1em\hfill\null\ \null\nobreak\hfill\kern3pt\vbox{\hrule\hbox
   {\vrule\kern1pt\vbox{\kern1.7pt\hbox{$\scriptscriptstyle{QED}$}
    \kern0.2pt}\kern1pt\vrule}\hrule}}
\def\END{\hskip0.1em\hfill\null\ \null\nobreak\hfill\kern3pt\vbox{\hrule\hbox
   {\vrule\kern1pt\vbox{\kern1.7pt\hbox{$\,\,\,\vspace{5pt}$}
    \kern0.2pt}\kern1pt\vrule}\hrule}}
\newcommand{\ie}{i.e$.$ }
\newcommand{\mto}{\mapsto}

\newcommand{\der}{\partial}

\DeclareMathOperator{\im}{im}

\DeclareMathOperator{\id}{id}
\DeclareMathOperator{\byd}{{\raisebox{.1ex}{:}{=}}}

\newcommand{\sub}{\subset}

\newcommand{\wed}{\wedge}
\newcommand{\com}{\!\circ\!}
\newcommand{\con}{\,\lrcorner\,}
\newcommand{\ten}{\!\otimes\!}

\newcommand{\ucar}[1]{\underset{#1}{\times}}

\newcommand{\owed}[1]{\overset{#1}{\wedge}}

\newcommand{\olin}[1]{\overline{#1}}

\newcommand{\alp}{\alpha}
\newcommand{\bet}{\beta}
\newcommand{\gam}{\gamma}
\newcommand{\del}{\delta}
\newcommand{\eps}{\epsilon}

\newcommand{\lam}{\lambda}
\newcommand{\sig}{\sigma}

\newcommand{\ome}{\omega}
\newcommand{\Gam}{\Gamma}

\newcommand{\Ome}{\Omega}

\newcommand{\vartht}{\vartheta}

\newcommand{\bg}{\boldsymbol{g}}

\newcommand{\bX}{\boldsymbol{X}}
\newcommand{\bY}{\boldsymbol{Y}}

\newcommand{\cC}{\mathcal{C}}
\newcommand{\cD}{\mathcal{D}}
\newcommand{\cE}{\mathcal{E}}

\newcommand{\bgam}{\boldsymbol{\gam}}
\newcommand{\bdel}{\boldsymbol{\del}}

\def\con{{\offinterlineskip\lower 1truept\hbox{\kern2pt
\vbox to7truept{\vfill\hbox to4truept{\hrulefill}}\vrule \kern3pt}}}
\newcommand{\D}{\mathcal{D}}

\newcommand{\myskip}{\vspace*{8pt}}
\newcommand{\For}[1]{\overset{#1}{\Lambda}}
\newcommand{\Con}[1]{\overset{#1}{\cal{C}}}
\newcommand{\Hor}[1]{\overset{#1}{\cal{H}}}
\newcommand{\Var}[1]{\overset{#1}{\cal{V}}}
\newcommand{\Thd}[1]{\overset{#1}{\Theta}}
\title{\textbf{On a class of polynomial Lagrangians}}
\author{Marcella Palese\thanks{Lecturer at the School. Partially
supported by GNFM of INDAM, MURST, University of Turin.}
\, and \,
Raffaele Vitolo\thanks{Partially
supported by GNFM of INDAM, MURST, Universities of Florence
and Lecce.}}
\date{}
\begin{document}

\maketitle

\begin{abstract}
\footnotesize{
In the framework of finite order variational sequences a new class of
Lagrangians arises, namely, \emph{special} Lagrangians.
These Lagrangians are the horizontalization of forms on a
jet space of lower order. We describe their properties together with
properties of related objects, such as Poincar\'e--Cartan and
Euler--Lagrange forms, momenta and momenta of generating forms,
a new geometric object arising in variational sequences. Finally, we
provide a simple but important example of special Lagrangian, namely
the Hilbert--Einstein Lagrangian.}

\myskip

\noindent \small{{\bf Key words}: Fibered manifolds, jet spaces,
variational sequences, polynomial Lagrangians.

}

\end{abstract}

\section{Introduction}

The theory of variational sequences provides a geometric framework for
the calculus of variations. In this theory the Euler--Lagrange operator
is just a morphism in an exact sequence of vector spaces (or sheaves
of vector spaces). Geometric objects like Lagrangians, momenta,
Poincar\'e--Cartan forms, Helmholtz conditions, find a nice
interpretation in the vector spaces of the sequence.

\myskip

We are concerned with some aspects of the theory of variational
sequences in finite order jet spaces (see \cite{Kup80,MaMo83a,Sau89}
for the basics on this subject), as was mainly developed in
\cite{Kru90,Kru93} (see also \cite{Vit97,Vit98} for further developments).
In this theory a subset of
$r$--th order Lagrangians is selected in a natural way by the geometric
structure of
finite order jets. Namely, this distinguished subset is made by
$r$--th order Lagrangians which are the horizontalization
\cite{Kru90,Vit97,Vit98} of $n$--forms on the jet space of order
$r-1$. Such Lagrangians are said to be \emph{special}.
The aim of this paper is to study in detail the
properties of special Lagrangians and related geometric objects.

\myskip

In the second section, we review the main results on the geometry of
spaces of jets $J_r\bY$ of a fibration $\bY \to\bX$.  We recall that
the tangent space $TJ_r\bY$ has a natural splitting when pulled back
to the higher order jet space $J_{r+1}\bY$.  Namely, it splits into the
(pullback of the) vertical subbundle plus a bundle which is
fiberwise isomorphic to $T\bX$.  Then, we introduce
horizontalization as the projection of forms on $J_r\bY$, or $r$--th
order forms, on forms on $J_{r+1}\bY$ having the highest exterior
factor $\owed{k}T^*\bX$ in their target space. We then recall
Krupka's theory of finite order variational sequences \cite{Kru90}.
A variational sequence on $J_r\bY$ is produced by taking the quotient
of de Rham sequence on $J_r\bY$ with respect to a sequence defined through the
kernel of horizontalization. The commutative diagram built by the
three sequences is said to be the (finite order) variational bicomplex.

In the third section, we describe the horizontalization of $k$--forms,
with $k\leq n$ (here $n$ is the dimension of the base manifold).
Horizontal $n$--forms of order $r+1$ are usually interpreted as
$(r+1)$--th order Lagrangians,
\cite{FeFr82,GaMu82,Gar74,Kol83,Kup80,MaMo83b}, but we prove that not
any horizontal form of order $r+1$ is the horizontalization of some
form of order $r$. We see that the components of horizontalized
$(r+1)$--th order $k$--forms have polynomial coefficients of degree
$k$ in the derivatives of order $r+1$.  Then, we define special
Lagrangians of order $r+1$ to be $n$--forms coming from the
horizontalization of a $r$--th order $n$--form.  We see that
horizontalization provides an isomorphism of the quotient space of
$n$--forms in the variational sequence on $J_r\bY$ with the space of
$(r+1)$--th order special Lagrangians.

The fourth section is devoted to Euler--Lagrange forms. We recall that
Euler--Lagrange forms are representatives of classes of $(n+1)$--forms
in the variational sequence \cite{Vit98}, through horizontalization
and a geometric version of Green's formula \cite{Kol83}.
In particular, we are able to split any horizontalized $(n+1)$--form,
which we call generating form, into an Euler--Lagrange form (not
necessarily induced by a Lagrangian) and the horizontal
differential (\ie the total divergence) of a form, which is said to
be a momentum for the generating form. These momenta were first
introduced in \cite{Vit98}, but here we study their properties in
detail. Then, we prove that it is possible to compute the
Euler--Lagrange form for special Lagrangians both in the standard way
and by using the commutativity of the variational bicomplex. Finally,
we describe the polynomial structure of the Euler--Lagrange forms
induced by special Lagrangians.

The fifth section contains a description of properties of momenta of
generating forms and their relationship with standard momenta of
(special) Lagrangians. We give a detailed analysis of their
uniqueness properties. Namely, we prove that such momenta are
uniquely determined either for $\dim\bX = 1$ or for generating forms
of order $2$. We show that such a momentum can be naturally determined
for generating forms of order $3$.
We think that momenta for generating forms could play an important role
in \emph{multisymplectic theories} (see
\cite{Hra99,Kan98} and their rich bibliography). These theories are a
generalization of symplectic formalism to field theory. They all
involve a closed $(n+1)$--form $\Ome$ on $J_1\bY$ as the main geometric
object. An analysis of these theories with the powerful tool of variational
sequences has never been attempted. Indeed, field equations can be
easily recovered via the Euler--Lagrange form induced by the
generating form $h(\Ome)$. Here, momentum should play an essential
role. This will be the subject of further studies. This is also a
good motivation for introducing and studying such objects.

In the sixth section, we give a characterization of Poincar\'e--Cartan
forms for both special and general Lagrangians. Namely, we prove that
a form $\theta$ is a Poincar\'e--Cartan form for a given Lagrangian if the
Lagrangian is the horizontalization of $\theta$, the vertical part
of $\theta$ is in the space of momenta and the momentum of the
generating form $h(d\theta)$ can be chosen to be zero. Of course, this can
also be taken as a definition of Poincar\'e--Cartan form inspired by
the variational sequences.

In the last section, we will show a relevant example of special
Lagrangian, namely the Hilbert--Einstein Lagrangian. We provide also
the related objects, such as the Poincar\'e--Cartan form, the momentum,
the Euler--Lagrange form and the momentum of the natural generating form.

\smallskip

Here, manifolds and maps between manifolds are assumed to be $\cC^{\infty}$.

\section{Jet spaces and variational sequences}\label{2}

In this section we recall some basic facts about jet spaces
\cite{Fe84,MaMo83a,Sau89} and Krupka's formulation of the finite
order variational sequence \cite{Kru90,Vit98}.

Our framework is a fibered manifold $\pi : \bY \to \bX$,
with $\dim \bX = n$ and $\dim \bY = n+m$.

For $r \geq 0$ we are concerned with the $r$--jet space $J_r\bY$;
in particular, we set $J_0\bY \equiv \bY$. We recall the natural fiberings
$\pi^r_s : J_r\bY \to J_s\bY$, $r \geq s$, $\pi^r : J_r\bY \to \bX$, and,
among these, the {\em affine\/} fiberings $\pi^r_{r-1}$.
We denote by $V\bY$ the vector subbundle of the tangent
bundle $T\bY$ of vectors on $\bY$ which are vertical with respect
to the fibering $\pi$.

Charts on $\bY$ adapted to $\pi$ are denoted by $(x^\lam ,y^i)$.
Greek indices $\lam ,\mu ,\dots$ run from $1$ to $n$ and they label
base coordinates, while Latin indices $i,j,\dots$ run from $1$ to $m$
and label fiber coordinates, unless otherwise specified.  We denote by
$(\der_\lam ,\der_i)$ and $(d^\lam, d^i)$ the local bases of vector
fields and $1$--forms on $\bY$ induced by an adapted chart,
respectively.

We denote multi--indices of dimension $n$ by the boldface Greek
letters $\bgam, \bdel$. We have
$\bgam = (\gam_1, \dots, \gam_n)$ with $0 \leq \gam_\mu$,
$\mu=1,\ldots,n$; by an abuse
of notation, we denote by $\lam$ the multi--index such that
$\bgam_{\mu}=0$ if $\mu\neq \lam$, $\bgam_{\mu}= 1$ if
$\mu=\lam$.
We also set $|\bgam| \byd \gam_{1} + \dots + \gam_{n}$ and $\bgam ! \byd
\gam_{1}! \dots \gam_{n}!$.

The charts induced on $J_r\bY$ are denoted by $(x^\lam,y^i_{\bgam})$, with $0
\leq |\bgam| \leq r$; in particular, we set $y^i_{\bf{0}}
\equiv y^i$. The local vector fields and forms of $J_r\bY$ induced by
the above coordinates are denoted by $(\der^{\bgam}_i)$ and
$(d^i_{\bgam})$, respectively.

In the theory of variational sequences a fundamental role is played by the
{\em contact maps\/} on jet spaces (see \cite{Fe84,MaMo83a,Sau89}).
Namely, for $r\geq 1$, we consider the natural complementary fibered
morphisms over $J_r\bY \to J_{r-1}\bY$
\beq
\D : J_r\bY \ucar{\bX} T\bX \to TJ_{r-1}\bY \,,
\qquad \qquad\quad
\vartht : J_r\bY \ucar{J_{r-1}\bY} TJ_{r-1}\bY \to VJ_{r-1}\bY \,,
\eeq
with coordinate expressions, for $0 \leq |\bgam| \leq r-1$, given by
\beq
\D = d^\lam\ten {\D}_\lam = d^\lam\ten
(\der_\lam + y^j_{\bgam+\lam}\der_j^{\bgam}) \,,
\qquad
\vartht = \vartht^j_{\bgam}\ten\der_j^{\bgam} =
(d^j_{\bgam}-y^j_{{\bgam}+\lam}d^\lam)
\ten\der_j^{\bgam} \,.
\eeq

We have
\bEq
\label{jet connection}
J_r\bY\ucar{J_{r-1}\bY}T^*J_{r-1}\bY =\left(
J_r\bY\ucar{J_{r-1}\bY}T^*\bX\right) \oplus\Con{*}_{r-1}[\bY]\,,
\eEq
where
$\Con{*}_{r-1}[\bY] \byd \im \vartht_r^*$.

Now, we introduce some distinguished sheaves of forms on jet
spaces \cite{Vit98}. Let $k \geq 0$.

\begin{enumerate}
\item
For $r \geq 0$, we consider the standard sheaf $\For{k}_r$
of $k$--{\em forms\/} on $J_r\bY$. We have the coordinate expression
\beq
\alp = \alp
{_{i_1 \dots i_{h} }^{\bgam_1 \dots \bgam_{h}}}
{_{\lam_{h+1} \dots \lam_{k}}} \,
d^{i_1}_{\bgam_1}\wed\dots\wed d^{i_{h}}_{\bgam_{h}}\wed
d^{\lam_{h+1}} \wed\dots\wed d^{\lam_{k}}\,.
\eeq
\item
For $0 \leq s \leq r$, we consider the sheaves $\Hor{k}_{(r,s)}$ and
$\Hor{k}_r$ of {\em horizontal forms\/}, \ie of local fibered morphisms over
$J_r\bY \to J_s\bY$ and $J_r\bY \to \bX$ of the type, respectively,
\beq
\alp : J_r\bY \to \owed{k}T^*J_s\bY
\qquad \textstyle{and} \qquad
\lam : J_r\bY \to \owed{k}T^*\bX \,;
\eeq
in coordinates $\lam =
\lam_{\lam_1\dots\lam_k}d^{\lam_1}\wed \dots\wed d^{\lam_k}$.
\item
Furthermore, we consider the subsheaf $\Hor{k}{_r^P} \sub \Hor{k}_{r}$
of local fibered morphisms $\alp \in \Hor{k}_r$ such that $\alp$ is a
{\em polynomial} fibered morphism over $J_{r-1}\bY \to\bX$ of degree
$k$. In coordinates, the components $\lam_{\lam_1\dots\lam_n}$ are
polynomials in $y^i_{\bgam}$ of degree $k$, where $|\bgam |= r$.
\item
For $ 0 \leq s < r$, we consider the subsheaf $\Con{k}_{(r,s)}
\sub \Hor{k}_{(r,s)}$ of {\em contact forms\/}, \ie
of local fibered morphisms over $J_r\bY \to J_{s}\bY$ of the type
\beq
\alp : J_r\bY\to \owed{k} \Con{*}_{s}[\bY]
\sub \owed{k}T^*J_{s}\bY \,,
\eeq
and the subsheaf $\Con{k}{_r} \sub \Con{k}_{(r+1,r)}$ of local
fibered morphisms $\alp \in \Con{k}_{(r+1,r)}$ such

such that $\alp = \owed{p}\vartht_{r+1}^* \com \Tilde{\alp}$, where
$\Tilde{\alp}$ is a section of the fibration $J_{r+1}\bY \ucar{J_r\bY}$
$\owed{p}V^*J_r\bY$ $\to J_{r+1}\bY$ which projects down onto
$J_{r}\bY$.

\end{enumerate}

The fibered splitting \eqref{jet connection} yields the
sheaf splitting
\bEq\label{graded}
\Hor{k}_{(r+1,r)}=\oplus_{l=0}^k
\Con{k-l}_{(r+1,r)}\wed\Hor{l}_{r+1}
\eEq
\cite{Sau89,Vit98}.  We set $h$ to be the restriction to $\For{k}_{r}$
of the projection of the above splitting on the term with the
highest degree of the horizontal factor.  We set also $v$ to be the
complementary projection $v \byd \id -h$.  We say $h$ to be the
\emph{horizontalization} of forms on jet spaces.

The splitting \eqref{jet connection} induces also a decomposition of
the exterior differential on $\bY$, $(\pi^{r+1}_r)^*\com \,d = d_H +
d_V$, where $d_H$ and $d_V$ are defined to be the {\em horizontal\/}
and the {\em vertical differential\/} \cite{Sau89}.

\myskip

We recall now Krupka's variational sequence on finite order jet
spaces \cite{Kru90}.

Let us denote by $\olin{d\ker h}$ the sheaf generated by the presheaf
$d\ker h$ (see \cite{Wel80}).  We set $\Thd{*}_{r}$ $\byd$ $\ker h$
$+$ $\olin{d\ker h}$.  In \cite{Kru90} it is proved that the following
diagram is commutative and that its rows and columns are exact:
\newdiagramgrid{Krupka}
{1,.5,1,.5,1,.7,1,1,1,.8,.8,1,1.2,.8,1,.7,.5,.5,1}
{.9,.9,.9,.9,.9,.9,.9,.9}
\beq
\diagramstyle[size=2.3em]
\begin{diagram}[grid=Krupka]
&& 0 && 0 && 0 && 0 &&&& 0 && 0  &&&&
\\
&& \dTo && \dTo && \dTo && \dTo &&&& \dTo && \dTo &&&&
\\
0 & \rTo & 0 & \rTo & 0 & \rTo &
\Thd{1}_r & \rTo^d & \Thd{2}_r & \rTo^d & \dots &
\rTo^d & \Thd{I}_r & \rTo^d & 0 & \rTo & \dots & \rTo & 0
\\
&& \dTo && \dTo && \dTo && \dTo &&&& \dTo && \dTo &&&&
\\
0 & \rTo & \R & \rTo & \For{0}_r & \rTo^d &
\For{1}_r & \rTo^d & \For{2}_r & \rTo^d & \dots & \rTo^d &
\For{I}_r & \rTo^d & \For{I+1}_r & \rTo^d & \dots & \rTo^d & 0
\\
&& \dTo && \dTo && \dTo && \dTo &&&& \dTo && \dTo &&&&
\\
0 & \rTo & \R & \rTo & \For{0}_r & \rTo^{\cE_{0}} &
\For{1}_r/\Thd{1}_r & \rTo^{\cE_{1}} & \For{2}_r/\Thd{2}_r & \rTo^{\cE_{2}} &
\dots & \rTo^{\cE_{I-1}} & \For{I}_r/\Thd{I}_r & \rTo^{\cE_{I}} &
\For{I+1}_r & \rTo^{d} & \dots & \rTo^{d} & 0
\\
&& \dTo && \dTo && \dTo && \dTo &&&& \dTo && \dTo &&&&
\\
&& 0 && 0 && 0 && 0 &&&& 0 && 0  &&&&
\end{diagram}
\eeq

The top row of the above diagram is said to be the $r$--th order
{\em contact sequence} and the bottom row is said to be the $r$--th order
{\em variational sequence\/} associated with the fibered manifold
$\bY\to\bX$ (see \cite{Kru90,Vit98} for the relationship with
calculus of variations).

The variational sequence can be read through some intrinsic isomorphisms of
quotient sheaves with sheaves of forms on jets \cite{Vit98}.
This shows the connection of the variational sequence with the
geometric formulations of the calculus of variations
\cite{FeFr82,GaMu82,Gar74,Kol83,Kup80,MaMo83b}. Here, we are concerned
with the columns of $n$ and $n+1$ forms.
\section{Special Lagrangians}

In this section, we introduce special Lagrangians as distinguished
representatives of equivalence classes in $\For{n}_r/\Thd{n}_r$. More
precisely, this representative will be obtained through
horizontalization.

\myskip

For $k\leq n$, let us set
\beq
\Hor{k}{_{r+1}^{h}} \byd h (\For{k}_r)\,.
\eeq
We say $\Hor{k}{_{r+1}^{h}}$ to be the sheaf of \emph{special
horizontal forms} of order $r+1$.

Special horizontal $k$--forms are $k$--th degree polynomial in higher order
derivatives, \ie $\Hor{k}{_{r+1}^{h}}\sub\Hor{k}{_{r+1}^P}$.
In fact, if $\alp\in\For{k}_r$, then
\beq
h(\alp) =
y^{i_1}_{\bgam_{1}+\lam_{1}} \dots y^{i_{h}}_{\bgam_{h}+\lam_{h}}
\alp
{_{i_1 \dots i_{h} }^{\bgam_1 \dots \bgam_{h}}}
{_{\lam_{h+1} \dots \lam_{k}}}
d^{\lam_{1}} \wed\dots\wed d^{\lam_{k}} \,,
\eeq
with $0 \leq h \leq k$.

\bRm\label{characterization}
The sheaf $\Hor{k}{_{r+1}^{h}}$ admits the following characterization
\cite{Vit98}: a section $\alp \in \Hor{k}{_{r+1}^P}$ is a section of
the subsheaf $\Hor{k}{_{r+1}^h}$ if and only if there exists a section
$\bet \in \For{k}_{r}$ such that
\beq
(j_r \sig)^{*}\bet = (j_{r+1}\sig)^{*}\alp
\eeq
for each section $\sig:\bX\to\bY$.\END
\eRm

If $\dim\bX = 1$ then the inclusion $\Hor{k}{_{r+1}^{h}} \sub
\Hor{k}{_{r+1}^P}$ is an equality. In fact, in this case the above
coordinate expression turns out to be the general coordinate
expression for a section of $\Hor{1}{^P_{r+1}}$.

If $\dim\bX \neq 1$, then the inclusion $\Hor{k}{_{r+1}^{h}} \sub
\Hor{k}{_{r+1}^P}$ \emph{is not} an equality, in general, due to the
above characterization.  We can check it via the following example.
Consider a $1$--form $\bet \in \For{1}_{0}$.  Then we have the
coordinate expressions $\bet = \bet_{\lam}d^\lam + \bet_{i}d^i$,
$h(\bet) = (\bet_{\lam} + y^i_\lam\bet_{i}) d^\lam$.  If $\alp \in
\Hor{1}{_{1}^P}$, then we have the coordinate expression $\alp =
(\alp_{\lam} + y^i_{\mu}\alp^{\mu}_{i}{_{\lam}}) d^\lam$.  It is
evident that, in general, there does not exist $\bet \in \For{1}_{r}$
such that $h(\bet) = \alp$.

\myskip

Let us recall that, according to the standard definition, an $r$--th
order \emph{Lagrangian} is defined to be a form $\lam\in\Hor{n}_r$
\cite{FeFr82,GaMu82,Gar74,Kol83,Kup80,MaMo83b}.

The horizontalization induces a natural sheaf isomorphism between
$\For{n}_r/\Thd{n}_r$ and $\Hor{n}{_{r+1}^h}$.  This motivates the
following definition.

\bDf
We say forms in $\Hor{n}{_{r+1}^h}$ to be \emph{special
Lagrangians} of order $r+1$.\END
\eDf

We also say a Lagrangian $\lam\in \Hor{n}_r$ to be \emph{general} if it is
not special. Equivalently, $\lam$ is general
either if it is not the horizontalization of a form in $\For{n}_{r-1}$,
or if $\lam \not\in\Hor{n}{_r^h}$.

\bRm\label{feature 1}
Special Lagrangians of order $r+1$ differs from both general
and polynomial Lagrangians of order $r+1$ for one essential feature:
they come from a form in $\For{n}_r$ through horizontalization.\END
\eRm

\section{Euler--Lagrange forms and special Lagrangians}

Here we describe the properties of Euler--Lagrange forms induced by
special Lagrangians.  We see that any Euler--Lagrange form (even not
induced by a Lagrangian) is obtained from a horizontalized
$(n+1)$--form by adding a suitable form which is an exact horizontal
differential.  The horizontalized $(n+1)$--form is said to be a
generating form, while a (horizontal) potential of the exact form is
said to be a momentum for the Euler--Lagrange form.  Then, we prove
that it is possible to compute the Euler--Lagrange form for special
Lagrangians both in the standard way and by using the commutativity of
the variational bicomplex.  Finally, we describe the structure of
Euler--Lagrange forms of special Lagrangians.

\myskip

The horizontalization induces the natural injective sheaf morphism
\beq
\left(\For{n+1}_{r}/\Thd{n+1}_{r}\right) \to
\left(\Con{1}{_{r}}\wed\Hor{n}{_{r+1}^h}\right) \big / h(\olin{d\ker h})
: [\alp ]\mto [h(\alp )]\,.
\eeq
Then, it can be proved that $h(\olin{d\ker h}) \sub
\olin{d_H(\Con{1}{_{r}}\wed\Hor{n-1}{_{r+1}^h})}$ \cite{Vit98}.
So, we can use Kol\'a\v r's geometric version of Green's integration by
part formula
to provide an isomorphism of the above quotient sheaf with a
sheaf of forms on jet spaces. Namely,
let us consider $h(\alp)\in\Con{1}_{r}\wed\Hor{n}{_{r+1}^h}$; such a
form is said to be a {\em generating form}.
It is proved in \cite{Kol83} that for any generating form $h(\alp)$
then there is a unique pair of sheaf morphisms
\bEq\label{Kolar}
E_{h(\alp)} \in \Con{1}_{(2r,0)}\wed\Hor{n}{_{2r+1}^h} \,,
\qquad
F_{h(\alp)} \in \Con{1}_{(2r,r-1)} \wed \Hor{n}{_{2r}^h} \,,
\eEq
such that $h(\alp)=E_{h(\alp)}+F_{h(\alp)}$ and
$F_{h(\alp)}$ is locally of the form $F_{h(\alp)} = d_{H}p_{h(\alp)}$,
with $p_{h(\alp)}\in \Con{1}_{(2r-1,r-1)}\wed\Hor{n-1}{_{2r}^h}$.
Note that a \emph{global} section $p_{h(\alp)}$ such that
$F_{h(\alp)} = d_{H}p_{h(\alp)}$ always exists
\cite{Fe84,FeFr82,GaMu82,Kol83}, essentially due to the fact that
$d_H$ has zero cohomology when restricted on certain subsequences (see
\cite{Al99} for a deeper discussion).

\bDf
Let $\alp\in\For{n+1}_r$. Then any form $p_{h(\alp)}$ is said to
be a {\em momentum} of the generating form $h(\alp)$.\END
\eDf

Notice that we are able to consider momentum also for
Euler--Lagrange forms which are not variational, \ie which do not
come from any Lagrangian.

\bRm
We think that momenta of this kind could play an important role
in the study of \emph{multisymplectic theories} (see
\cite{Hra99,Kan98} and their rich bibliography). These theories are a
generalization of symplectic formalism to field theory and all of them
involve a closed $(n+1)$--form $\Ome$ on $J_1\bY$ as the main geometric
object. An analysis of these theories with the powerful tool of variational
sequences has never been attempted. Indeed, field equations can be
easily recovered via the Euler--Lagrange form induced by the
generating form $h(\Ome)$. Here, momentum could play an essential
role.\END
\eRm

The above yields the sheaf isomorphism
\bEq\label{Euler iso}
\left(\Con{1}{_{r}}\wed\Hor{n}{_{r+1}^h}\right) \big / h(\olin{d\ker h})
\to \Var{n+1}_r : [h(\alp)]\mto E_{h(\alp)}\,,
\eEq
where $\Var{n+1}_{r} \byd
\left(\Con{1}{_{r}}\wed\Hor{n}{_{r+1}^h} +
d_H (\Con{1}_{(2r,r-1)}\wed\Hor{n-1}_{2r}) \right) \cap
\left(\Con{1}_{(2r+1,0)}\wed\Hor{n}_{2r+1}\right)$ \cite{Vit98}.
It is now clear that generating forms of order $r+1$ provide all
Euler--Lagrange forms in the quotient space of $(n+1)$--forms in the
variational sequence of order $r$.

\myskip

Let us recall the standard definition of Euler--Lagrange form
and momentum for a Lagrangian $\lam\in\Hor{n}_r$
\cite{Fe84,FeFr82,GaMu82,Kol83,Kup80}.
We apply \eqref{Kolar} to obtain
$d\lam = E_{d\lam}+d_Hp_{d\lam}$ for any choice of $p_{d\lam}$. We say

-- $E_{d\lam}$ to be the \emph{Euler--Lagrange form} of the Lagrangian
$\lam$;

-- $p_{d\lam}$ to be a {\em momentum} of the Lagrangian $\lam$.

The momentum of a Lagrangian is uniquely defined only in some special cases
\cite{Fe84,FeFr82,GaMu82,Kol83}. Namely, either if $\dim\bX = 1$ or
if $r = 1$. If $r=2$ then we are able to naturally determine
$p_{d\lam}$ through a further assumption \cite{Kol83}. If $r=3$ then
there does not exist, in the general situation, a natural
$p_{d\lam}$ \cite{Kol93}. Anyway, an intrinsic choice of $p_{d\lam}$ is
always possible \cite{Kol83}.

We show that the operator $\cE_n$ of the variational sequence
associates to any Lagrangian its Euler--Lagrange form
through the above isomorphism \eqref{Euler iso}.

\bPr\label{Euler}
Let $\lam\in\Hor{n}{_{r+1}^h}$ and $\bet \in \For{n}_r$ such that
$h(\bet) = \lam$. Then we have $\cE_n(\lam) = E_{h(d\bet)}$.
Moreover, $E_{h(d\bet)} = E_{d\lam}$ .
\ePr
\bPf
By the above decomposition formula,
$h(d\bet) = E_{h(d\bet)}+d_Hp_{h(d\bet)}$ for any choice of $p_{h(d\bet)}$.
But the commutativity of the diagram
\begin{diagram}
\For{n}_r & \rTo^d & \For{n+1}_r
\\
\dTo & & \dTo
\\
\Hor{n}{_r^h} & \rTo^{\cE_n} & \Var{n+1}_r
\end{diagram}
yields $\cE_n(\lam) = E_{h(d\bet)}$.
As for the second result, we consider $\lam$ as being a form $\lam \in
\For{n}_{r+1}$. In this case, $\cE_n(\lam) = E_{d\lam}$.
By the inclusion of the $r$--th variational bicomplex into the
$(r+1)$--th one \cite{Kru90,Vit98}, we obtain $E_{h(d\bet)} =
E_{d\lam}$.\QED
\ePf

If $\lam\in\Hor{n}_r$ is general, then the form $E_{d\lam}$
is defined on $J_{2r}\bY$, and has a peculiar
structure with respect to the derivative coordinates of order greater than
$r$. In fact, if we assign to the variables $y^i_{\bgam}$ with
$| \bgam | = r+s$ the weight $s$, then it is easily seen that $E_{d\lam}$
is a polynomial with weighted degree $r$ with respect to $y^i_{\bgam}$, with
$| \bgam | > r$ \cite{KoMo90}.

\bCr
If $\lam\in\Hor{n}_r$ is special, then the form $E_{d\lam}$ is defined
on $J_{2r-1}\bY$, and the coefficients of the polynomials in
$E_{d\lam}$ are polynomials of (standard) degree $n+1$ with respect to
the coordinates $y^i_{\bgam}$, with $|\bgam | = r+1$.
\eCr
\section{Momentum and special Lagrangians}

Now, we describe general properties of momentum for generating
forms $h(\alp) \in\Con{1}_{r}\wed\Hor{n}{_{r+1}^h}$. Then, we see the
relationship with momenta for special Lagrangians.

We recall the coordinate expression $h(\alp)=
\tilde{\alp}^{\bgam}_i\vartheta^i_{\bgam}\wed\ome$, where
$\tilde{\alp}^{\bgam}_i$
are polynomials of (standard) degree $n +1$ with respect to
the coordinates $y^i_{{\bgam}}$, with $| {\bgam} | = r+1$, with
coefficients the components of $\alp$.

\myskip

As we already said, \emph{global momenta} $p_{h(\alp)}$ for
any generating form $h(\alp)$ always exist.
This is essentially due to the fact that $d_H$ has zero cohomology.
A proof of this can be found in an early work by Kol\'a\v r (see
references in \cite{Kol83}). See also \cite{Al99} for a
cohomological proof.

Then, we check uniqueness properties of $p_{h(\alp)}$.
Of course, if $\dim\bX = 1$ then $p_{h(\alp)}$ is unique. This is because
$d_H\,p_{h(\alp)}=0$ implies $p_{h(\alp)}=0$, as it is easily seen in
coordinates.

\bRm
There exists a natural sheaf morphism \cite{Kol93,MaMo83b,Sau89,Vit98}
\beq
p: \Con{1}_{(r,1)}\wed\Hor{n}{_{r}} \to \Con{1}_{(r,0)}\wed
\Hor{n-1}{_{r}} \,.
\eeq
If $\phi\in\Con{1}_{(r,1)}\wed\Hor{n}{_{r}}$ has the coordinate
expression $\phi = \phi_{i} \, \vartheta^i\wed \ome +
\phi_{i}^{\lam} \, \vartheta^i_{\lam}\wed\ome$,
then we have the coordinate expression $
p_{\phi} = \phi_{i}^{\lam} \, \vartht^i\wed \ome_{\lam}$.\END
\eRm

\bTh\label{uniqueness I} (Uniqueness I).
Let $\alp\in\For{n}{_1}$. Then, the momentum $p_{h(\alp)}$ of $h(\alp)$
is unique. We have the coordinate expression
\beq
p_{h(\alp)} = \tilde{\alp}^\lam_i \vartht^i\wed\ome_\lam\,.
\eeq
\eTh
\bPf
In fact, we deduce the above coordinate expression from \eqref{Kolar}.
Then, it is clear that
$p_{h(\alp)}$ is defined up a $n$--form whose horizontal
differential vanish. It is easy to see in coordinates that such a
form must be zero.\QED
\ePf

\bRm
It is easy to verify that if we start with $\alp\in\For{n}{_2}$
we obtain $h(\alp) \in \Con{1}_{(3,1)}\wed\Hor{n}{_{3}^h}$, so
$h(\alp)$ \emph{is not} in the domain of $p$.\END
\eRm

In the case $r = 2$ there is not a unique choice of momentum for
the generating form $h(\alp)$. But we are able to choose it in a natural way.

\bRm
There exists a natural sheaf morphism \cite{Kol93,Vit98}
\beq
s: \Con{1}_{(r,1)}\wed\Hor{n-1}{_{r}} \to \Con{1}_{(r,0)}\wed
\Hor{n-2}{_{r}} \,.
\eeq
If $p\in\Con{1}_{(r,1)}\wed\Hor{n-1}{_{r}}$ has the coordinate
expression $p = p_{i}^{\;\;\mu} \, \vartheta^i\wed \ome_\mu +
p_{i}^{\lam\mu} \,\vartheta^i_{\lam}\wed\ome_\mu$, then we have the
coordinate expression $s(p) = p_{i}^{\lam\mu} \, \vartht^i\wed
\ome_{\lam\mu}$.\END
\eRm

\bTh (Uniqueness II).
Let $\alp\in\For{n}{_2}$. Then, there exists a unique
momentum $p_{h(\alp)}$ of $h(\alp)$ such that $s(p_{h(\alp)}) = 0$.
We have the coordinate expression
\beq
p_{h(\alp)} =
(\tilde{\alp}_{i}^{\lam} - D_{\mu}\tilde{\alp}{_{i}^{\mu +\lam}}) \,
\vartht^i\wed \ome_{\lam} +
\tilde{\alp}{_{i}^{\mu +\lam}} \, \vartht^i_{\mu}\wed \ome_{\lam}\,.
\eeq
\eTh
\bPf
Suppose that $p_{h(d\bet)} = p_i^\lam \vartheta^i\wed\ome_\lam +
p_i^{\lam\mu} \vartheta^i_\mu\wed\ome_\lam$. Then
$s(p_{h(\alp)}) = 0$ yields $p_i^{\lam\mu}=-p_i^{\mu\lam}$.
By \eqref{Kolar} one obtains the above $p_{h(d\bet)}$
as the unique momentum fulfilling the above requirement.\QED
\ePf

\bRm
It is easy to verify that if we start with $h(\alp)\in\For{n}_3$
then we obtain $h(\alp) \in
\Con{1}_{(4,2)}\wed\Hor{n}{_{4}^h}$, hence $p_{h(\alp)} \in
\Con{1}_{(5,2)}\wed\Hor{n}{_{5}^h}$, so that $p_{h(\alp)}$ \emph{is not}
in the domain of $s$.\END
\eRm

\bRm
The reader could have realized that the above proofs go in the same
way as in the case of general Lagrangians $\lam$ \cite{Kol83}.
The difference is that here we used generating forms $h(\alp)$ instead.
This means that, even if results refer to orders $1$ and $2$ as in
the case of Lagrangians, generating forms are of order $2$ and $3$,
respectively.\END
\eRm

\myskip

Now, we deal with the interplay between the two kind of
momenta that we introduced: momenta of (special) Lagrangians and
momenta of generating forms. Let $\lam \in
\Hor{n}{_{r+1}^h}$ be a special Lagrangian.  Then, there exists
$\bet\in\For{n}_r$ such that $h(\bet) = \lam$. So, we can consider
the generating form $h(d\bet)$ and evaluate its momentum $p_{h(d\bet)}$.
It is natural to ask the relationship between the momentum
$p_{d\lam}$ of $\lam$ and the momentum $p_{h(d\bet)}$ of $h(d\bet)$.

First of all, we note that $\bet$ is not unique, hence all uniqueness
results referring to $p_{h(d\bet)}$ that we evaluated above cannot be
related to $\lam$.

\bTh
We have $h(d\bet)=h(d_H\,v(\bet)) + d\lam$, hence the momenta
$p_{h(d\bet)}$ and $p_{d\lam}$ can be chosen to be equal if and only if
$h(d_H\,v(\bet)) = 0$.
\eTh
\bPf
In fact,
\begin{align*}
h(d\bet)& = h((d_H+d_V)(\lam + v(\bet)))
\\
& = h(d_H\,v(\bet) + d_V\lam + d_V\,v(\bet))
\\
& = h(d_H\,v(\bet)) + d_V\lam \,,
\end{align*}
where, in this case, $d_V\lam = d\lam$.\END
\ePf

\bCr
Let $\lam\in \Hor{n}_r\sub\Hor{n}{_{r+1}^h}$ be a general Lagrangian. Then,
the momenta $p_{h(d\bet)}$ and $p_{d\lam}$ can be chosen to be equal.
\eCr
\bPf
In fact, in this case $\bet = \lam$ hence $v(\bet) = 0$ and we can
choose $p_{h(d\bet)} = p_{d\lam}$.\QED
\ePf
\section{Poincar\'e--Cartan forms and special Lagrangians}

Here, we give a characterization of Poincar\'e--Cartan forms in the
framework of variational sequences. This characterization is inspired
by and formulated through special $(r+1)$--th order Lagrangians, but
obviously it holds also for general Lagrangians of any order.

\myskip

We recall that, given a Lagrangian $\lam\in\Hor{n}_r$, we define the
form $\theta_\lam \byd \lam + p_{d\lam} \in \For{n}_{2r-1}$ to be a
\emph{Poincar\'e--Cartan form}
\cite{Fe84,FeFr82,GaMu82,Kol83,Kup80,MaMo83b,Sau89}. Such a
definition is motivated by the fact that the differential of the
Poincar\'e--Cartan form splits into the sum of the Euler--Lagrange form
for $\lam$ plus a contact form, namely
$d\theta_\lam = E_{d\lam}+d_Vp_{d\lam}$. Uniqueness consideration for
the Poincar\'e--Cartan form are the same as momentum \eqref{Kolar}.

Our characterization of Poincar\'e--Cartan forms is inspired by the
fact that we can choose zero momentum for the
generating form $h(d\theta_\lam)$!

\bTh
Let $\lam\in\Hor{n}{_{r+1}^h}$ be a special Lagrangian. Then there
exists a unique class of forms $\theta \in \For{n}_{2r}$ fulfilling

1 -- $h(\theta) = \lam$;

2 -- $v(\theta) \in \Con{1}_{2r}\wed\Hor{n-1}_{2r}$;

3 -- $h(d\theta) = E_{h(d\theta)}$, or we can choose zero momentum for the
generating form $h(d\theta)$.

Namely, $\theta = \theta_\lam$.
\eTh
\bPf
In fact, requirements 1 and 2 imply that $\theta$ should be of the
form $\theta = \lam + p$, with $p\in\Con{1}_{2r}\wed\Hor{n-1}_{2r}$.
Now,
\beq
h(d\theta) = h(d_Hp) + d_V\lam = h(d_Hp) + E_{d\lam} - d_Hp_{d\lam}
\eeq
But $h(d\theta) = E_{h(d\theta)} = E_{d\lam}$ due to theorem \ref{Euler}.
Moreover, requirement 2 imply $h(d_Hp) = d_Hp$. Summing up,
$d_H(p-p_{d\lam}) = 0$, hence $p$ is also a momentum for
\nolinebreak $\lam$.

Conversely, it is trivial to see that Poincar\'e--Cartan forms
fulfill the requirements of the theorem.\QED
\ePf

\bRm
We would like to justify the requirements of the above theorem.
The first requirement is obviously necessary. The second requirement
is a requirement of `minimality' of the vertical part of $\theta$ with
respect to the splitting \ref{graded}. The third requirement is
inspired by the main property of Poincar\'e--Cartan forms that we
recalled at the beginning of the section.
\END
\eRm

\bRm
Of course, these requirements could be taken as a
definition of Poincar\'e--Cartan form naturally provided by variational
sequences. This in the same spirit as definitions of Lagrangians,
Euler--Lagrange forms and momenta in the above framework.

Moreover, we stress that the structure of the variational sequence, via theorem
\ref{Euler}, characterizes the Poincar\'e--Cartan form as a Lepagean
equivalent of
$\lam$ (see e.g. \cite{Kru93,Kru99}).
The last requirement of the above theorem explicitely expresses that the
generating form
of the Poincar\'e--Cartan form coincides with the associated
Euler--Lagrange form.
\END
\eRm

\section{The Hilbert--Einstein Lagrangian}\label{example}

In this brief section we show an important and simple example of
special Lagrangian, namely the Hilbert--Einstein Lagrangian.
We also derive all related geometric objects like the momentum of the
Hilbert--Einstein Lagrangian, its Euler--Lagrange
form and the momentum of the Euler--Lagrange form.

\myskip

Let $\dim\bX = 4$ and $\bX$ be orientable. Let
$Lor(\bX)$ be the bundle of Lorenzian metrics on $\bX$ (provided
that it exists). Local fibered coordinates on $J_{2}(Lor(\bX))$ are
$(x^{\lam}; g_{\mu\nu}, g_{\mu\nu,\sig}, g_{\mu\nu,\sig\rho})$.

The Hilbert--Einstein Lagrangian is the form $\lam_{HE} \in
\Hor{4}_{2}$ defined by $\lam_{HE} = L_{HE}\ome$, were
$L_{HE}=r\,\sqrt{\bg}$. Here $r: J_{2}(Lor(\bX)) \to \R$ is the function
such that, for any Lorenz metric $g$, we have $r\circ j_2g = s$,
being $s$ the scalar curvature associated with $g$, and $\bg$ is the
determinant of $g$.

The function $L_{HE}$ is a \emph{linear} function in the second derivatives
of $g$. In fact, let us set $G^{\alp\bet\eps\gam} \byd
g^{\alp\eps}g^{\bet\gam} + g^{\alp\gam}g^{\bet\eps}
- 2 g^{\alp\bet}g^{\eps\gam}$; then we have \cite{Fr88}
\beq
r = \frac12 G^{\alp\bet\eps\gam} \left(g_{\eps\gam ,\alp\bet} +
g_{\mu\nu}\Gam_{\alp\bet}^\mu\Gam_{\eps\gam}^\nu\right)\,.
\eeq
We can prove even more. Indeed, $\lam_{HE} \in \Hor{4}{_{2}^{h}}$. In fact,
the momentum for the second order Lagrangian
$\lam_{HE}$ (in the sense of \cite{Kol83}) turns out to be \cite{Fr88}
\begin{align*}
p_{\lam_{HE}} &= \frac12 \left(G^{\alp\bet\eps\gam}
g_{\mu\nu} \der^{\mu\nu ,\lam}
\left(\Gam_{\alp\bet}^\mu\Gam_{\eps\gam}^\nu\right) -
\cD_\rho (G^{\lam\rho\mu\nu} \sqrt{\bg})\right)\;
\vartheta_{\mu\nu}\wed\ome_\lam +
\\
& \hphantom{=} \frac12 G^{\lam\rho\mu\nu}\;
\sqrt{\bg} \vartheta_{\mu\nu ,\rho}\wed\ome_\lam\,,
\end{align*}
and the Poincar\'e--Cartan form
\begin{align*}
\theta_{\lam_{HE}} & =
\frac12 G^{\alp\bet\eps\gam}
g_{\mu\nu}\Gam_{\alp\bet}^\mu\Gam_{\eps\gam}^\nu\; \sqrt{\bg}\,\ome +
\\
& \hphantom{=} \frac12 \left(G^{\alp\bet\eps\gam}
g_{\mu\nu} \der^{\mu\nu ,\lam}
\left(\Gam_{\alp\bet}^\mu\Gam_{\eps\gam}^\nu\right) -
\cD_\rho (G^{\lam\rho\mu\nu} \sqrt{\bg})\right)\;
\vartheta_{\mu\nu}\wed\ome_\lam +
\\
& \hphantom{=} \frac12 G^{\lam\rho\mu\nu}\;
\sqrt{\bg} \vartheta_{\mu\nu ,\rho}\wed\ome_\lam\,.
\end{align*}
Of course, $\theta_{\lam_{HE}} \in \For{4}_1$. Moreover, a direct
computation shows that
\beq
h(\theta_{\lam_{HE}}) = \lam_{HE}\,.
\eeq
So, \emph{$\lam_{HE}$ is a special Lagrangian ($r=1$).}

In view of the previous results, its Euler--Lagrange
form should be an element $E_{d\lam_{HE}} \in
\Con{1}_{(2,0)}\wed\Hor{4}_3$. But, due to a property of $\lam_{HE}$
\cite{Fr88}, we have $E_{d\lam_{HE}} \in
\Con{1}_{(2,0)}\wed\Hor{4}_2$. Of course, a direct computation shows
that $E_{d\lam_{HE}} = G \byd R - \frac12\,s\,g$, $R$ being the Ricci
tensor of the metric $g$.

Another important consideration is that we can also compute
$E_{d\lam_{HE}}$ through proposition \ref{Euler}, namely as
$E_{d\lam_{HE}} = E_{h(d\bet)}$. In this case,
we have a natural candidate of $\bet$, namely we can take $\bet =
\theta_{\lam_{HE}} \in \For{4}_1$. So,
\beq
d\theta_{\lam_{HE}} = E_{d\lam_{HE}}+d_Vp_{d\lam_{HE}}
\eeq
(see the above section), which yields the natural generating form
$h(d\theta_{\lam_{HE}}) = E_{d\lam_{HE}} = E_{h(d\theta_{\lam_{HE}})}$.
So, by theorem \ref{uniqueness I}, the unique momentum of the
generating form $h(d\theta_{\lam_{HE}})$ is the zero form. This very peculiar
behaviour is due to the geometric structure of general relativity. It
is also an example of a special Lagrangian with a non trivial momentum
and whose momentum of the natural generating form vanishes.

\subsection*{Acknowledgments}

Thanks are due to Prof. I. Kol\'a\v r for useful
discussions. The first author would like also to thank the Head of
the Winter School Geometry and
Physics, Prof. J. Van\v zura, and the whole Organizing Committee, for the
nice and stimulating stay in $Srn\acute{\imath}$.

\noindent Commutative diagrams have been drawn by Paul Taylor's
\texttt{diagrams} macro package.


\newpage

\noindent {\em Authors' addresses}:

\medskip

\noindent{\footnotesize Marcella Palese}\\
{\footnotesize Department of Mathematics, University of Torino}
\\{\footnotesize Via C. Alberto 10, 10123 Torino, Italy}
\\{\footnotesize E--mail: \email{palese@dm.unito.it}} \\

\medskip

\noindent{\footnotesize Raffaele Vitolo}\\ 
{\footnotesize Department of Mathematics ``E. De Giorgi",
University of Lecce}
\\{\footnotesize Via Arnesano, 73100 Lecce, Italy}
\\{\footnotesize E--mail: \email{Raffaele.Vitolo@unile.it}}

\end{document}